\documentclass[manuscript]{aastex}

\slugcomment{Not to appear in Nonlearned J., 45.}

\shorttitle{Hydrogen Fluoride in High-Mass Star-forming Regions}
\shortauthors{Emprechtinger et al.}

\begin{document}


\title{Hydrogen Fluoride in High-Mass Star-forming Regions}


\author{M. Emprechtinger\altaffilmark{}}
\affil{California Institute of Technology, Cahill Center for Astronomy and Astrophysics 301-17, Pasadena, CA 91125, USA}
\email{emprecht@caltech.edu}

\author{R. R. Monje\altaffilmark{}}
\affil{California Institute of Technology, Cahill Center for Astronomy and Astrophysics 301-17, Pasadena, CA 91125, USA}

\author{F. F. S.. van der Tak\altaffilmark{}}
\affil{SRON Netherlands Institute for Space Research, Groningen, NL}

\author{M. H. D. van der Wiel\altaffilmark{}}
\affil{Institute for Space Imaging Science, Department of Physics and Astronomy, University of Lethbridge, Lethbridge, AB, T1K 3M4, Canada}
\affil{Kapteyn Astronomical Institute, University of Groningen, Groningen, NL}
\affil{SRON Netherlands Institute for Space Research, Groningen, NL}

\author{D. C. Lis\altaffilmark{}}
\affil{California Institute of Technology, Cahill Center for Astronomy and Astrophysics 301-17, Pasadena, CA 91125, USA}

\author{D. Neufeld\altaffilmark{}}
\affil{Johns Hopkins University, Baltimore MD,  USA}

\author{T. G. Phillips\altaffilmark{}}
\affil{California Institute of Technology, Cahill Center for Astronomy and Astrophysics 301-17, Pasadena, CA 91125, USA}

\author{C. Ceccarelli\altaffilmark{}}
\affil{UJF-Grenoble 1/CNRS-INSU, Institut de Plan\'étologie et d'Astrophysique de Grenoble (IPAG) UMR 5274, Grenoble, France}




\begin{abstract} 
Hydrogen fluoride has been established to be an excellent tracer of molecular hydrogen in diffuse clouds. In denser environments, however, the HF abundance has been shown to be approximately two orders of magnitude lower. We present Herschel/HIFI observations of HF~$J=1-0$ toward two high-mass star formation sites, NGC6334~I and AFGL~2591. In NGC6334~I the HF line is seen in absorption in foreground clouds and the source itself, while in AFGL~2591 HF is partially in emission.  We find an HF abundance with respect to H$_2$ of $1.5\cdot 10^{-8}$ in the diffuse foreground clouds, whereas in the denser parts of NGC6334~I, we derive a lower limit on the HF abundance of $5\cdot 10^{-10}$. Lower HF abundances in dense clouds are most likely caused by freeze out of HF molecules onto dust grains in high-density gas. In AFGL 2591, the view of the hot core is obstructed by absorption in the massive outflow, in which HF is also very abundant ($3.6\cdot 10^{-8}$) due to the desorption by sputtering.  These observations provide further evidence that the chemistry of interstellar fluorine is controlled by freeze out onto gas grains. 
\end{abstract}






\section{Introduction}

Hydrogen fluoride (HF) is an exceptional molecule, because of its peculiar chemistry. Fluorine is one of the few atoms which has a greater affinity to hydrogen than hydrogen itself, and thus the reaction
$$
H_2 + F \rightarrow HF+H
$$
is highly exothermic ($\rm \Delta E\approx 16000$~K, although the activation of this reaction energy is 400~K and quantum mechanical tunneling through this barrier has to be considered). The destruction of HF occurs by the relatively slow photo-dissociation ($\rm1.17\cdot10^{-10}~s^{-1}$) and reactions with ions, such as He$^+$, H$_3^+$, and C$^+$, which are rare compared to H$_2$.  Thus HF formation is much more efficient than its destruction under most interstellar conditions. As a consequence HF is considered to be the main reservoir of fluorine in molecular regions, i.e, where $\rm H_2/H>1$, and therefore it is thought to be an ideal tracer of H$_2$. Astrochemical models predict an abundance of HF relative to H$_2$ (all fractional abundances given in this paper are relative to H$_2$) of $3.6\cdot10^{-8}$ in such clouds (\citealt{NWS05}).

The first detection of HF was reported by \cite{NZS97}, who observed the HF~$J=2-1$ transition ($\nu=2463.43$~GHz) with relatively low resolution (R=9600) in absorption towards Sgr~B2, using the {\it{Infrared Space Observatory (ISO)}}. However, because high densities or strong radiation fields are needed to populate the HF~$J=1$ level, the HF column density derived from this observation is subject to large uncertainties.  With the advent of the {\it{HIFI}} instrument \citep{dHP10} aboard the \emph{Herschel Space Observatory} \citep{PRP10}, high resolution spectroscopy ($\rm R>10^6$) of the ground state rotational transition of HF has become possible for the first time. Several HF observations in diffuse cloud along the lines of sight toward strong sub-millimeter dust continuum sources, such as W31C (\citealt{NSP10}), W49N and W51 (\citealt{SNP10}), and Sgr~B2(M) (\citealt{MEP11}) have been reported. HF~$J=1-0$ is detected in absorption toward all of these continuum sources, which was expected considering the extreme conditions necessary to populate the excited rotational states of this molecule. 
These observations prove that HF is an excellent tracer of H$_2$ in diffuse clouds, although the abundances of HF relative to H$_2$, derived by comparison with CH, are $\sim 1.5\cdot10^{-8}$, with an uncertainty of 46\%, and thus a factor of two below the chemical model predictions. Furthermore, HF turned out to be a much more sensitive tracer of H$_2$ than the commonly used CO, which is in agreement with the model prediction that HF is more abundant than CO in diffuse clouds with low extinction (\citealt{NWS05}). \cite{SNP10}, for example, detected a diffuse cloud on the line of sight toward W51, not seen in CO.

The only HF abundance calculated in a high-mass star-forming region, and not in a diffuse foreground cloud, has been reported by \cite{PBL10}, who detected HF~$J=1-0$ in absorption towards Orion~KL. Orion~KL is a peculiar source, heated from the front, which shows hardly any lines in absorption at submillimeter wavelengths. The HF abundance derived by Phillips et al. is a lower limit of $1.6\cdot10^{-10}$, approximately two orders of magnitude lower than the HF abundances reported in diffuse clouds. This result can be either explained by geometrical effects, due to the incomplete coverage of the continuum source by absorbing material, or, under the assumption that most fluorine is bound in HF, is caused by HF molecules sticking onto the surfaces of dust grains under high-density conditions. The only three detections of HF in emission so far are in the immediate vicinity of the AGB-star IRC+10216 \citep{ACW11}, the Orion Bar \citep{vON12} and towards the Seyfert 1 Galaxy Mrk~231 \citep{vIM10}.

To investigate the abundances of HF in dense regions in more detail, we analyze here the HF~$J=1-0$ spectra towards two high-mass star-forming regions, NGC6334~I and AFGL~2591. 
NGC6334 is a relatively nearby (1.7 kpc; \citealt{N78}) high-mass star-forming region, which harbors sites of many stages of protostellar evolution (\citealt{SH89}). Single-dish continuum observations at sub-millimeter wavelength show that a total mass of 200~$M_\odot$ is associated with NGC6334 I (\citealt{S00}). The molecular hot core NGC6334~I, studied extensively over the last decades (e.g. \citealt{BWT08, BWT07}; \citealt{HBM06}), is associated with an ultra compact HII region (\citealt{dRD95}) and shows a very line-rich spectrum (\citealt{SCT06}, \citealt{TWM03}). {\it{HIFI}} observations in NGC6334~I have led to the detection of CH \citep{vvL10}, H$_2$O \citep{ELB10}, H$_2$O$^+$ (\citealt{OML10}), and H$_2$Cl$^+$ (\citealt{LPN10}), showing that hydrides are common toward this source.  Furthermore, a bipolar outflow (\citealt{LSP06}; \citealt{BWT08}) and H$_2$O, OH, CH$_3$OH class~II, and NH$_3$ masers have been detected (e.g., \citealt{KJ95}; \citealt{EvM96}; \citealt{WLT07}). \cite{RSW11} studied the radial structure of several high-mass star-forming cores, including NGC6334~I. This investigation, mainly based on HCN observations with the APEX telescope, revealed a density law of $n\propto r^{-1.5}$. The temperature also follows a power law, with an index of 0.83 and 0.4 in the inner and the outer part, respectively.  Interferometric data (SMA), however, revealed that the internal structure of NGC6334~I is much more complex. The hot core consists of four compact condensations within a 10$''$ region, emitting about 50\% of the continuum flux, whereas the other 50\% stem from the extended envelope \citep{HBM06}.

AFGL~2591 is an isolated high-mass star-forming region located within the Cygnus-X region. The distance toward AFGL~2591 was reported to be between 0.5~kpc and 2.0~kpc, and for modeling purposes a distance of 1~kpc was assumed (e.g., \citealt{vvE99}). However, new measurement base on VLBI parallax measurements of H$_2$ masers suggest that the distance might be as large as 3.3~kpc \citep{RBS12}. Based on a distance of 1~kpc, the luminosity of AFGL~2591 is $\rm\sim2\cdot10^4~L_{\odot}$, however, adopting the newly measured distances the luminosity would be approximately a factor ten higher. The embedded central star, which is obscured in the visible by the massive envelope, has an estimated mass of 16~M$_\odot$ and an effective temperature of 33,000~K \citep{vM05}. Besides the circumstellar envelope a massive outflow with an outflow velocity of $\sim 20$~km\,s$^{-1}$ with respect to the systemic velocity has been detected \citep{LTS84}. In addition, \cite{HM95} found a second, faster ($\rm v_{out}\approx 40 km\,s^{-1}$), but weaker outflow.  Many attempts to model the structure of AFGL~2591 have been carried out assuming spherical symmetry and a power-law density and temperature structure (e.g., \citealt{vvE99}; \citealt{vvE00}; \citealt{Dvv02}; \citealt{dHF09}). A cavity, caused by the outflow might by present (\citealt{vvE99}). \cite{vvS11} mapped AFGL~2591 in the frequency range from 330--373~GHz with the JCMT. In this frequency range they found $\sim 160$ lines, of which 35 show extended emission ($>15''$). These observations reveal a small scale structure ($\leq 10^4$~AU), and a line of sight velocity gradient is apparent in most molecules. Their modeling suggests that a non isotropic structure or a velocity gradient must be present on $\sim10^4$~AU scales in order to explain the observations.

\section{Observations}\label{obs}

The observations, used in this paper,  are part of the {\it{C}}hemical {\it{HE}}rschel {\it{S}}urveys of {\it{S}}tar forming regions (CHESS; \citealt{CBB10}), a Herschel guaranteed time key program, which has conducted unbiased spectral line surveys of several star-forming regions. The HF~$J=1-0$ spectra presented in this paper are observed in two sources: NGC6334~I ($\alpha_{2000}:~17^h20^m53.32^s,~\beta_{2000}: -35^\circ46'58.5''$) and AFGL~2591 ($\alpha_{2000}:~20^h29^m24.9^s,~\beta_{2000}: +40^\circ11'21.6''$). The observation identification (ObsID) number of the scans are 1342206594 and 1342196510 for the observations in NGC6334~I and AFGL~2591, respectively. Data were taken on May~12$^{th}$~2010 (AFGL~2591) and October~14$^{th}$~2010 (NGC6334~I).

Both observations were carried out with the {\it{HIFI}} instrument \citep{dHP10} on board of the {\it{Herschel Space Observatory}} (\citealt{PRP10}) using the double beam switch mode ($180''$ chopper throw), and have been reduced with the HIPE pipeline \citep{O10} version~5.1 and version~3.0, respectively. The double sideband (DSB) spectra were observed with a redundancy of eight, which allows the deconvolution and isolation of the single sideband (SSB) spectra \citep{CS02}. 
In the case of NGC6334~I the deconvolution was done in HIPE and the deconvolved SSB spectra were exported to the FITS format for subsequent analyses in the IRAM/GILDAS package \footnote{www.iram.fr/IRAMFR/GILDAS/}. The spectra of AFGL~2591 where exported to the FITS format right after the basic data reduction (level 2), and further data reduction and deconvolution was done using GILDAS. 
The spectra shown in this paper are equally weighted averages of the H and V polarization.

\begin{center}
\begin{figure}
\includegraphics[angle=-90,scale=.35]{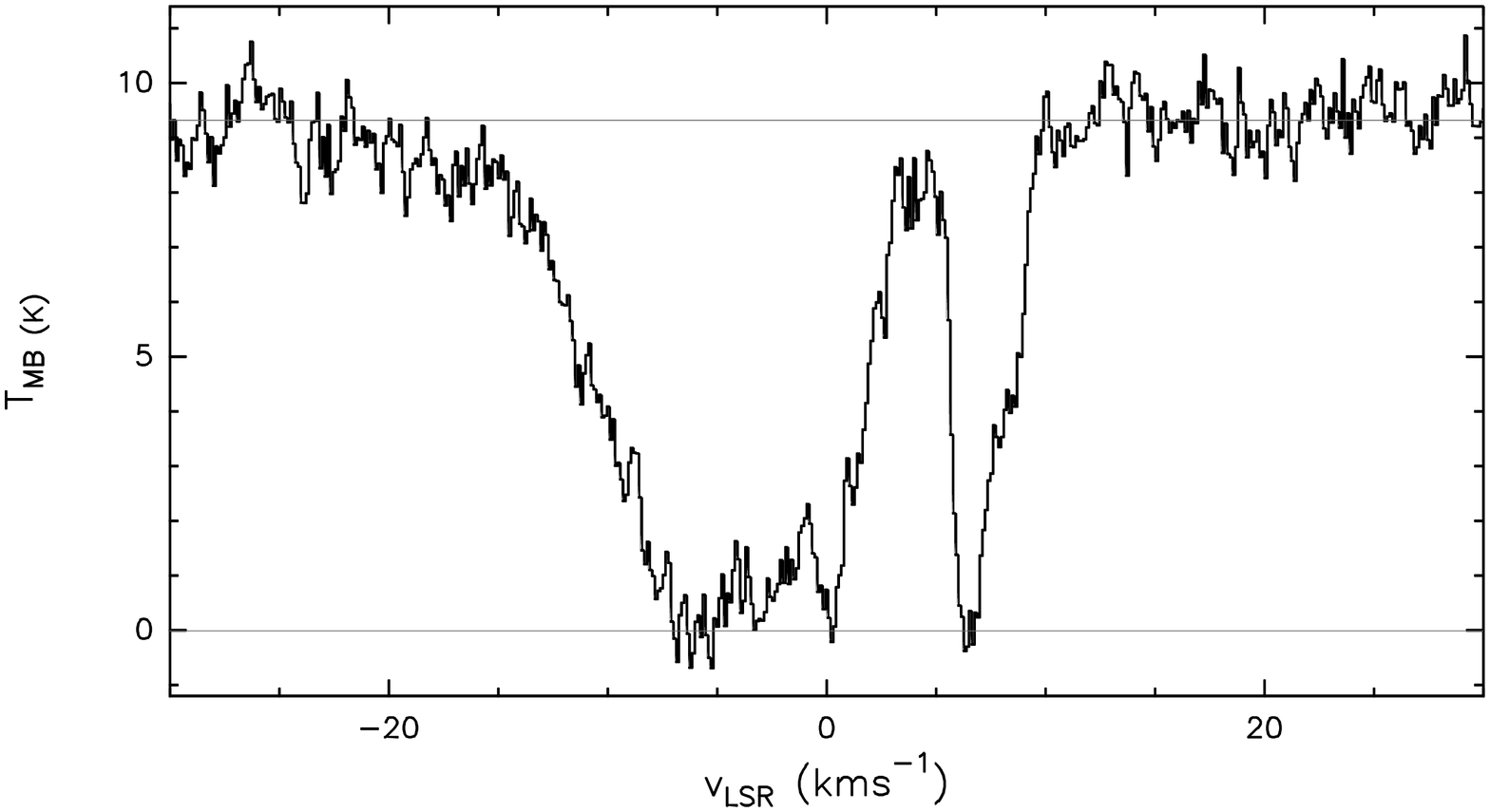}\includegraphics[angle=-90,scale=.35]{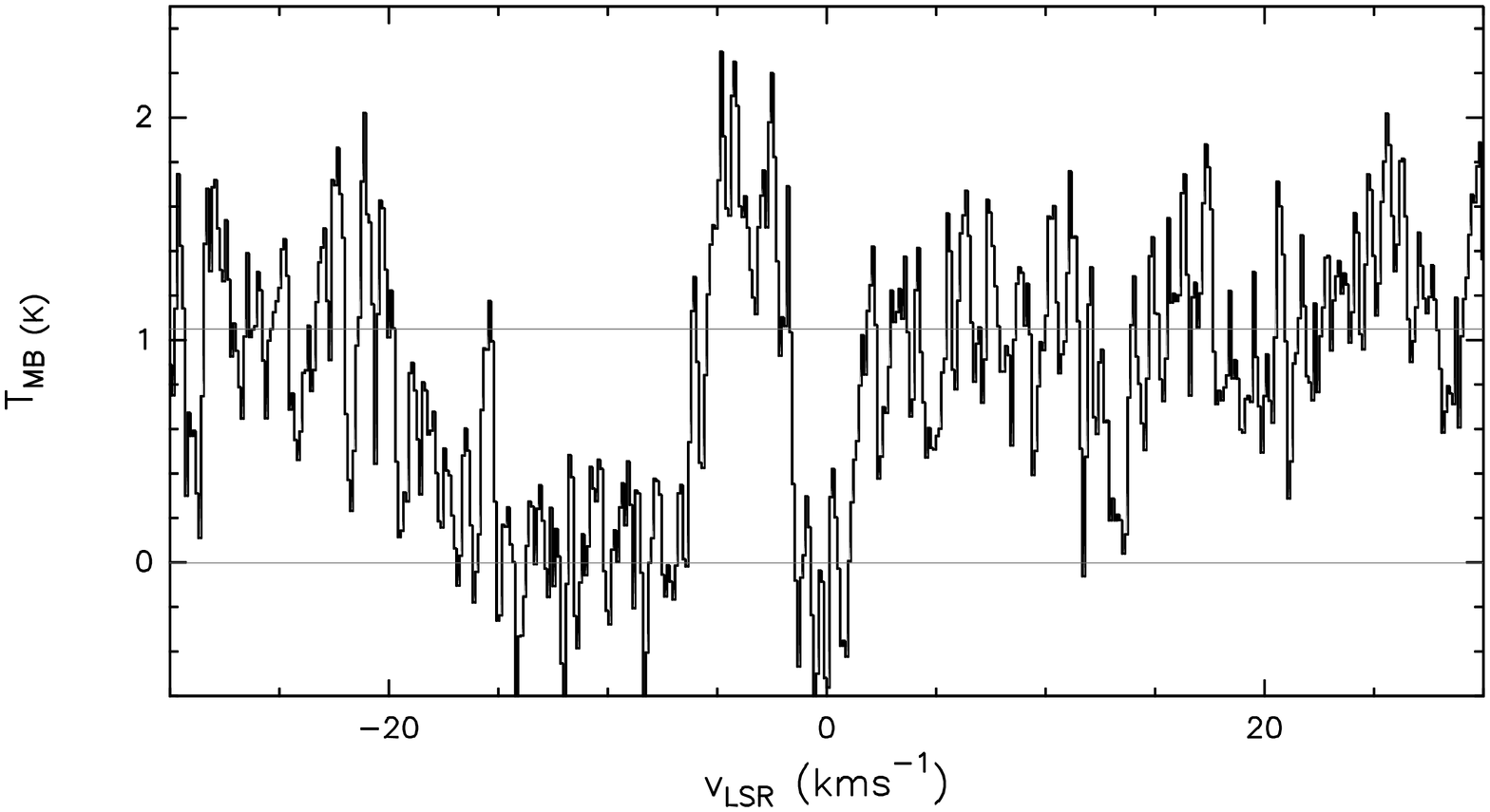}
\caption{HF~$J=1-0$ spectra in NGC6334~I (left) and AFGL~2591 (right). The respective continuum level as well as the zero level are marked with the gray lines.}\label{spectra}
\end{figure}
\end{center}

The HF~$J=1-0$ line ($\nu = 1232.47627$~GHz) has been clearly detected in both sources. The beam size of Herschel at this frequency is 18$''$ and a main beam efficiency of 0.62 was applied. Both spectra are displayed in Figure~\ref{spectra}.

\section{Results \& Analysis}\label{RaA}

As shown in Figure~\ref{spectra}, the HF line seen is in absorption, in both sources, over a wide range of velocities. Only in the velocity interval from $-$8~km\,s$^{-1}$ to $-$2~km\,s$^{-1}$ in AFGL~2591 a weak emission is seen. In many cases the HF absorption is completely saturated. Assuming that all HF molecules are in the rotational ground state, one can calculate from the absorption depth the line opacity and subsequently the HF column density using following formulae:
\begin{equation}
\tau = -ln\left(\frac{T_{\rm{MB}}}{T_{\rm{back}}}\right)
\label{taucold}
\end{equation}
where $T_{MB}$ is the measured brightness temperature and $T_{back}$  the temperature of the background continuum,
and
\begin{equation}
N_{HF}=\frac{8\pi^{3/2}\cdot \Delta v}{2 \sqrt{ln 2}~\lambda^3 \cdot A}\frac{g_l}{g_u}\cdot\tau
\label{Ntaucold}
\end{equation} 
where $\lambda$ is the wavelength (243.2~$\mu$m), $A$ the Einstein~A~coefficient ($\rm2.422\cdot10^{-2}~s^{-1}$), and $g_l$ and $g_u$ are the statistical weights of the lower and upper state (1 and 3, respectively). The assumption of all HF molecules being in the ground state is justified by the high Einstein~A~coefficient and the low rates for collisional excitation \citep{Gprep}, corresponding to a critical density of $\rm n_{crit}=1.9\cdot10^{8}~cm^{-3}$, and the relatively high energy of the $J=1$ state (50~K). As shown by Van der Tak~(submitted), most alternative, non-thermal excitation mechanisms, such as infrared pumping and formation pumping, play only a negligible role. 
Excitation caused by submillimeter radiation on the other hand is possible in the hot inner part of high mass star forming regions and has to be considered. We will discuss the potential influence of this mechanism on the HF excitation individually for each source later on.

\subsection{NGC6334~I}

\begin{center}
\begin{figure}
\includegraphics[angle=-90,scale=.65]{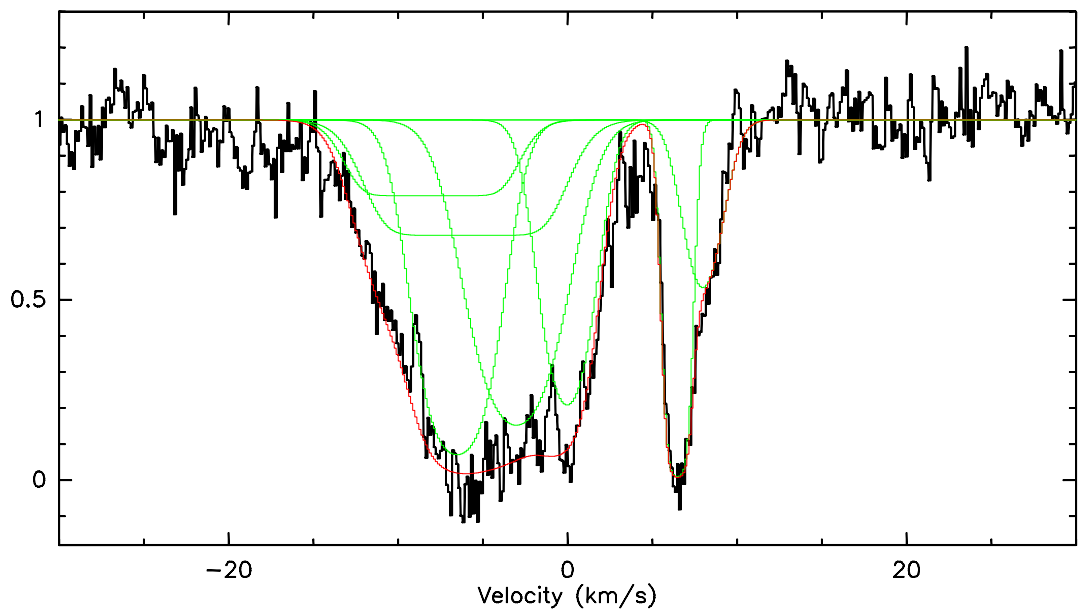}
\caption{Comparison of the normalized spectrum of HF (black) and the model (red). The green lines show the individual components.}\label{ngcpfit}
\end{figure}
\end{center}

Many of the features seen in the HF spectrum toward NGC6334~I originate from four foreground clouds, of which two reside in the immediate environment of NGC6334~I, whereas the two remaining clouds are not associated with the source itself. These foreground clouds, which have been identified in CH spectra \citep{vvL10} are at velocities of --3.0~km\,s$^{-1}$, 0.0~km\,s$^{-1}$, +6.5~km\,s$^{-1}$, and +8.0~km\,s$^{-1}$, whereas the velocities of the individual components of NGC6334~I itself are between $-6.0$ km\,s$^{-1}$ and $-8.0$ km\,s$^{-1}$. From fitting several C$^{18}$O lines observed with {\it{HIFI}}, we know that the source itself can be modeled by three components at $v_{\rm LSR}$=--6.0~km\,s$^{-1}$, --8.0~km\,s$^{-1}$, and --6.5~km\,s$^{-1}$ (Plume priv. communication). The first two components are identified as two of the embedded sub cores (SMA~1 and SMA~2 following \citealt{HBM06}), whereas the later component represents the lower density  envelope. Because, HF is a sensitive tracer of the molecular gas, we expect to see all these components in the HF spectrum as well. A massive outflow, detected in water (Emprechtinger et al.~submitted) and CO \citep{LSP06} in the velocity range between -100~km\,sand +60~km\,s, is not seen in HF, neither in emission nor in absorption. The upper limit of the optical depth of HF in the blue lobe is 0.88 what results in an upper limit on the HF column density of $8.42\cdot 10^{13}~\rm cm^{-2}$ and therefore in an upper limit on the HF abundance of $5\cdot 10^{-8}$, using the H$_2$ column density given by \cite{LSP06}. The upper limit of the integrated intensity of the red lobe is 2.69~K\,km\,s$^{-1}$, which results, assuming the canonical HF abundance, in an upper limit on the HF excitation temperature of 10~K. Both limits are consistent with standard HF abundances and negligible HF excitation. Thus the non-detection of the outflow in HF is another indication that most HF molecules are in the ground state in this source. The foreground components and the envelope are expected to be more extended than the continuum sources in NGC6334~I, and thus $T_{back}$ in equation~\ref{taucold} is equal to the measured continuum level in the spectra. The components associated with embedded cores, however, absorb only continuum emitted from the respective core. Furthermore, we cannot be certain that the absorbing material covers all the continuum radiation emitted by the core. Therefore, even at very high optical depths only a fraction of the background radiation is absorbed. From interferometric observations at $\lambda=1.3$~mm (\citealt{HBM06}) we estimate the contribution of each of the cores to be approximately 25\% of the total continuum flux. This 25\% value is only a rough estimate, since the actual value depends on the exact dust temperature and the coverage of the continuum source.  Therefore we use the actual continuum contribution of the cores as free fitting parameters $f_{c1}$ and $f_{c2}$. Using the centroid velocities and line widths of the seven components derived from the CH and C$^{18}$O observations, and assuming that all HF molecules are in the ground state, the only free parameters in addition to $f_{c1}$ and $f_{c2}$ are the HF column densities of the individual components. This approach is valid for all components except the +8.0~km\,s$^{-1}$ foreground cloud, which is narrower in HF ($\rm \Delta v=2.5~km\,s^{-1}$) than in CH ($\rm \Delta v=3.5~km\,s^{-1}$).
Using equations~\ref{taucold} \& \ref{Ntaucold} with the free parameter $f_{c1}$, $f_{c2}$ and the HF column densities as model to fit the observed spectra, we used the $X ^2$ method to determine the column densities listed in Table~\ref{ngcfit}; results are shown in Figure~\ref{ngcpfit}. The reduced $X^2$ for the best fit result is 1.07, but all results with a reduced $X^2$ lower than four have to be considered as possible solutions. Thus the HF column densities can be determined only with an  accuracy of 20\%-50\% for each individual component.

\begin{table}
\begin{center}
\caption{Fit results for the individual velocity components in NGC6334~I.\label{ngcfit}}
\begin{tabular}{lccc}
\hline\hline
Component & $v$ & $\Delta v$ & $N$ \\
& [km\,s$^{-1}$] & [km\,s$^{-1}$] & [cm$^{-2}$] \\ 
\hline
Foreground & $-3.0$ & 5.0 & $2.4\pm1.2\cdot10^{13}$ \\
Foreground & 0.0  & 3.0 & $1.5\pm0.3\cdot10^{13}$ \\
Foreground & +6.5  & 1.2 & $>1.4\cdot10^{13}$ \\
Foreground & +8.0  & 2.5 & $4.0\pm0.8\cdot10^{12}$ \\
Envelope Cloud  & $-6.5$ & 4.0 & $2.7\pm1.3\cdot10^{13}$ \\
SMA~1        & $-6.0$ & 6.0 & $2.0\pm1.0\cdot10^{14}$ \\
SMA~2        & $-8.0$ & 5.0 & $2.0\pm1.0\cdot10^{14}$ \\
\hline
f$_{c1}=0.32\pm0.06$ & & f$_{c2}=0.21\pm0.06$ & \\
\hline
\hline
\end{tabular}
\end{center}
\end{table}

The HF column density of the 0~km\,s$^{-1}$ and the +8.0~km\,s$^{-1}$ foreground components are well determined ($\pm 20\%$), whereas the $-$3.0~km\,s$^{-1}$ component is less constrained ($\pm 50\%$) due to the blending with the absorption features from NGC6334~I itself. For the +6.5~km\,s$^{-1}$ component only a lower limit for the column density can be given since this line is optically thick. However, the opacity of the +6.5~km\,s$^{-1}$ component is moderate, because we do not see any effects of extremely high optical depth, such as a line width wider than the optically thin CH line, or a saturated line shape. 
Two narrow residuals seen at $v_{\rm LSR}$=--1~km\,s$^{-1}$ and --9~km\,s$^{-1}$, are not reproduced by our fit. Considering that both residuals are located in the wings of major absorption features, a possible explanation is a velocity or line width gradient within individual components.
These two features may also be caused by emission lines from other species. However, there are no plausible candidates in the JPL database \footnote{http://spec.jpl.nasa.gov/home.html} or the CDMS catalog \citep{MSS05}
CH is known to be an excellent tracer of H$_2$ at densities below $\rm 10^4~cm^{-3}$ with an abundance of $3.5\cdot 10^{-8}$ with respect to H$_2$ (\citealt{SRF08}). Using the CH column densities reported by \cite{vvL10} to derive the H$_2$ column density, we are able to determine the HF abundance relative to H$_2$ in the diffuse foreground clouds (see Table~\ref{H2comp}).

\begin{table}
\begin{center}
\caption{Abundances of HF in the foreground components in NGC6334~I\label{H2comp}}
\begin{tabular}{lccc}
\hline\hline
vel. Comp & $N$(HF) & $N$(H$_2$) & $X$(HF)\\
 (km\,s$^{-1}$) & (cm$^{-2}$) & (cm$^{-2}$) & \\
\hline
$-$3.0   &    $2.40\cdot10^{13}$  & $1.8\cdot 10^{21}$ & $1.3\cdot10^{-8}$	\\
0.0    &    $1.50\cdot10^{13}$  & $6.0\cdot 10^{20}$ & $2.5\cdot10^{-8}$	\\
+6.5  &   $>1.4\cdot10^{13}$& $8.6\cdot 10^{20}$ & $>1.6\cdot10^{-8}$	\\
+8.0  &   $4.0\cdot10^{12}$  & $6.0\cdot 10^{20}$  & $6.7\cdot10^{-9}$	\\
\hline
\end{tabular}
\end{center}
\end{table}

The HF abundance in the clouds at --3.0~km\,s$^{-1}$, 0.0~km\,s$^{-1}$, and +6.5~km\,s$^{-1}$ agree fairly well with the abundances found in other diffuse clouds (\citealt{MEP11}, \citealt{SNP10})  However, the +8.0~km\,s$^{-1}$ component exhibits an HF abundance which is significantly lower. The +8.0~km\,s$^{-1}$ component is also the component in which the HF line is much narrower than CH. Furthermore, the abundance of CH$^+$ is significantly enhanced in this cloud compared to the other foreground components (Lis et al.~in prep.), indicating that the chemical and physical properties of the gas may be very different \citep{GFP09}.

Since observations of Orion~KL suggest that the abundance of HF in dense gas in high-mass star-forming regions is lower than in diffuse clouds (\citealt{PBL10}), we attempted to derive the HF abundance in the components associated with NGC6334~I itself. The fits for the SMA~2 core and the envelope material are relatively good and HF column densities can be determined with an accuracy of 50\%. The column density of the SMA~1 core, however, is poorly constrained since this component is blended with several other components. 
As can be seen in Figure~\ref{ngcpfit}, the HF absorption features originating from SMA~1 and SMA~2 are largely overlapping, which results in a large ambiguity of the individual continuum coverage factors ($\pm 0.2$). However, their sum can be defined quite well as $0.53\pm0.05$.
The fit results for SMA~1 presented in Table~\ref{ngcfit} are based on the assumption that the HF abundance relative to H$_2$ in SMA~1 and SMA~2, which have similar physical conditions, is the same.  To estimate the H$_2$ column density, we are using the C$^{18}$O column densities derived by Plume (priv. communication) from an LTE fit to seven C$^{18}$O transitions. We are assuming standard values for the C$^{16}$O/C$^{18}$O ratio (500) and the C$^{16}$O abundance ($9.5\cdot10^{-5}$). Furthermore we divide the resulting column density by two, because approximately half of the material emitting the CO lines resides behind the continuum sources, and therefore does not contribute to the HF absorption. The resulting HF abundances range from $3.7\cdot10^{-10}$ in the envelope to $5-8\cdot10^{-10}$ in the dense cores. 
These values are approximately two orders of magnitude lower than the abundances in diffuse clouds and predicted by models \citep{NWS05} but comparable to the lower limits found by \cite{PBL10} in Orion~KL.
In these calculations we did not consider submillimeter radiative excitation, and therefore these results are lower limits. However, taking the observed continuum values \citep{HBM06} and using LVG calculations, we found that we  underestimate the HF column density by less than a factor four for conditions present in the dense cores. Therefore, the low abundances of HF derived in NGC6334~I itself cannot be explained by HF excitation alone.

The calculated HF abundances are based on the assumption that approximately half of the mass associated with NGC6334~I resides in front of the corresponding continuum source. However, the actual structure of the source is not known, and thus the HF abundance might be underestimated. But, it should be noticed that it is very unlikely that the unknown geometry of NGC6334~I is the sole reason for the low HF abundance derived here. Assuming that only geometrical effects account for the low HF abundance would imply that in all three components (SMA~1, SMA~2, and the envelope cloud) as well as in Orion~KL, the bulk of the molecular gas ($\sim 98\%$) resides behind the main continuum source.

\subsection{AFGL~2591}

The spectrum of AFGL~2591 is much simpler than that of NGC6334~I; in addition to one foreground absorption component at $\sim 0$~km\,s$^{-1}$, only one broad component ranging from $-$6~km\,s$^{-1}$ to $-$20~km\,s$^{-1}$ can be seen. In the velocity range between the two absorption features a weak HF emission is detected. Both absorption features are most likely very optically thick, since all background radiation is completely absorbed in the corresponding velocity ranges. Unfortunately, the continuum  of AFGL~2591 is much weaker than the one of NGC6334~I and thus the signal to noise ratio is much lower, resulting in a lower limit for $\tau$ of 1.6.

The profile of the HF~$J=1-0$ line is very peculiar and only the profiles of the water ground-state transitions p-H$_2$O~$1_{11}-0_{00}$ and p-H$_2$O~$2_{12}-1_{01}$ look somewhat similar, although the emission component is much stronger in the water lines \citep{Cvv11}.  The broad absorption component is clearly blue shifted with respect to the systemic velocity of --5.7~km\,s$^{-1}$ of AFGL~2591, and thus most likely caused by the massive outflow. This outflow is also seen in CO and HCO$^+$ (\citealt{LTS84}; \citealt{HM95}), and centers at a velocity of --7~km\,s$^{-1}$, slightly blueshifted with respect to the envelope \citep{v11}. He also derived an H$_2$ column density of the outflow of $\rm 1.5\cdot10^{22}~cm^{-2}$, based on observations of several CO line and radiative transfer calculations. This H$_2$ column density leads to an HF column density of $\rm 2~\cdot10^{14}~cm^{-2}$ assuming an HF abundance of $1.5\cdot10^{-8}$.  Using a line width of 10~km\,s$^{-1}$, which is the approximate velocity width of the outflow, this HF column density corresponds to a $\tau$ of $\sim 9$ (using equation~\ref{Ntaucold}), which is in agreement with the high optical depth seen in the observed spectrum. HF abundances of the order of $\sim 3\cdot10^{-10}$, as found in star-forming regions, are clearly too low to cause the observed deep absorption. This is a very interesting aspect, because the temperature (100--150~K) and density ($\rm 10^{5}-10^{6}~cm^{-3}$) are comparable to the conditions in high-mass star-forming regions. It should be noted that a shock with a velocity of $\rm\approx 15~km\, s^{-1}$, as derived form the spectrum, may cause a partial dissociation of H$_2$ (up to 80\%) due to energetic collisions with neutral particles \citep{FP10}. This may cause an underestimation of the total hydrogen column density in the outflow (2N(H$_2$)+N(H)). However, collisions with enough energy to destroy H$_2$ can potentially destroy HF and CO, too. Furthermore, in the absence of H$_2$, the HF production is hindered. Therefore, we are able to derive the HF abundance relative to H$_2$, but not relative to the total hydrogen density. The error of the HF abundance relative to H$_2$ is determined by the uncertainty of the H$_2$ column density derived from CO. We estimate that these errors are definitely less than the degree of the dissociation of H$_2$ and therefore clearly lower than one order of magnitude.

The foreground absorption line width (equivalent width is 3.2~km\,s$^{-1}$) is clearly larger than the corresponding feature in the CO lines ($\rm\Delta v=0.8~km\,s^{-1}$; \citealt{v11}), and more comparable to the width of the feature seen in CH ($\rm\Delta v=1.8~km\,s^{-1}$, \citealt{BBv10}). Using the line width of CH as the $\Delta v$ caused by Doppler broadening for HF, and fitting the absorption feature using equations~(\ref{taucold}) and (\ref{Ntaucold}), leads to a minimum optical depth of eight, and thus a minimum HF column density of $\rm4\cdot10^{13}~cm^{-2}$ in the foreground cloud. Comparing this HF column density with the H$_2$ column density found by \cite{v11} based on CO observations, yields an HF abundance relative to H$_2$ of $>1.3\cdot10^{-8}$, which is a typical value found in diffuse clouds and close to that found in the clouds toward NGC6334~I. However, the CO abundance in diffuse clouds might be significantly lower than the canonical value of $10^{-4}$. \cite{SWT07} found CO abundances with respect to H$_2$ ranging from $2.75\cdot10^{-7}$ to $2.75\cdot10^{-5}$ in material on the line of sight toward several stars. Hence the total H$_2$ column density toward AFGL~2591 may be underestimated and, because we derive only lower limits on the HF column density, the uncertainty of the derived HF abundance is very large.

To get a better understanding of the HF spectrum in AFGL~2591, especially of the outflow and the HF emission, we model this source using RATRAN \citep{Hv00}. RATRAN is a Monte Carlo radiative transfer code, which includes excitation by collision of the molecules with H$_2$ and excitation by submillimeter radiation.
It calculates a grid of spectra integrated along a pencil beam, which are subsequently convolved to the desired spatial resolution.  
The Einstein coefficients of the modeled transitions are taken from the LAMDA database \citep{Svv05}, whereas newly calculated collision rates are used \citep{Gprep}.
The source model is based on the ``static model'' of \cite{vvS11}, who modeled the envelope of AFGL~2591 assuming a power law density distribution ($n=r^{-1.0}$) and used a self-consistent solution for the dust temperature \citep{vvE00}. To take the outflow component into account, we add an expansion velocity to each shell in the model of \cite{vvS11}, which increases with radius as
$$
v_{exp} = v_{\infty}\cdot\left(1-\frac{r_0}{r}\right)^\beta
$$ 
where $v_{\infty}$ is the maximum velocity of the outflow and $r_0$ is the radius at which the outflow starts. $\beta$, which was set to two, is a measure of how quickly the maximum velocity is reached; $r_0$ was set to a value so that the H$_2$ column density outside this radius is comparable to $\rm 1.5\cdot10^{22}~cm^{-2}$, the N$_{H_2}$ found in the outflow \citep{v11}.  At radii smaller than $r_0$ the expansion velocity is zero.  The maximum velocity is set to 9~km\,s$^{-1}$ to match the width of the absorption feature.  The modeled HF spectrum is not very sensitive to the actual velocity profile, and a proper solution can be found for profiles with moderately increasing $v_{exp}$. The foreground component is fitted using the foreground option in RATRAN assuming an optical depth of eight and a line width of 1.8~km\,s$^{-1}$.

\begin{figure}
\includegraphics[angle=0,scale=.50]{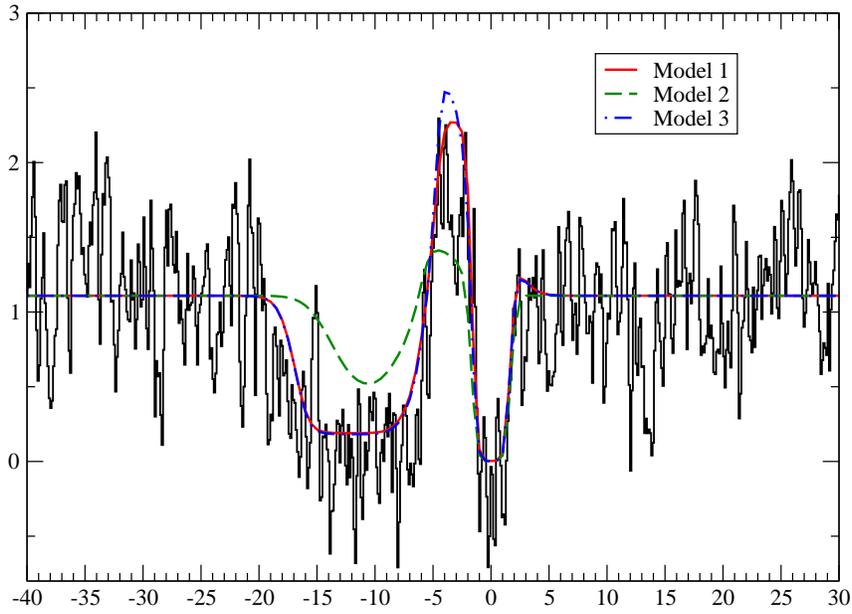}
\caption{Comparison of the observed HF~$J=1-0$ transition (black) with radiative transfer models. Model~1 (red): HF abundance  is $1.5\cdot 10^{-8}$; Model~2 (green): HF abundance is $5\cdot 10^{-10}$; Model~3 (blue): HF abundance is $5\cdot 10^{-10}$ in the quiescent gas and  $3.6\cdot 10^{-8}$  in the outflow\label{HAm}.} 
\end{figure}

Deriving the HF abundance is somewhat difficult, because the absorption features are optically thick. Hence we calculate models with three different abundances relative to H$_2$. In Model~1 we assume an abundance of $1.5\cdot 10^{-8}$, the HF abundance found in diffuse clouds, whereas in Model~2 an HF abundance of $5\cdot10^{-10}$ as found in the dense part of AFGL~2591 is adopted. In Model~3 we assume a value of $5\cdot10^{-10}$ in the inner part of the source, where $v_{exp}\rm=0~km\,s^{-1}$ and $1.5\cdot10^{-8}$ in the outer part. 
A comparison of the three models with observations is shown in Figure~\ref{HAm}. Models~1 and 3 fit the observations quite well, whereas an HF abundance as low as $5\cdot10^{-10}$ in the outflow can be ruled out. The model calculations show as well that the emission seen at $v_{\rm LSR}\rm\approx -4~kms^{-1}$, stems from the red lobe on the far side of the continuum source, and not from the hot core itself.
The emission at $v\rm _{LSR}\approx -4$~km\,s$^{-1}$ is overestimated by both, Model~1 and Model~3. This may be due to geometry effects, since the outflow is modeled by a shell, whereas it is in fact a collimated outflow. 
The absorption of all radiation in the blue lobe of the outflow indicates that most HF molecules are in the $J=0$ state, and thus excitation by collisions with H$_2$ or any other effective excitation mechanism can be ruled out. As suggested by our model calculation,  the emission from the red lobe is most likely caused by absorption and immediate re-emission of 1.2~THz photons from the continuum source of AFGL~2591.

Because none of the features of the HF spectra stems from the hot quiescent part of AFGL~2591, we cannot distinguish whether the abundance in the inner part is similar to the diffuse cloud abundance or to the lower limits found in the high-mass star-forming region NGC6334~I and Orion~KL \citep{PBL10}. The assumption of a uniform, low HF abundance in the dense parts of high-mass star-forming regions is consistent with our data.

\section{Summary \& Discussion}\label{DIS}

We determine the abundance of HF in very different physical environments, and obtain very different results. In agreement with previous studies, we find the HF abundance of $\sim 1.5\cdot10^{-8}$ in diffuse clouds, implying that about half of the available fluorine is in the form of HF in the gas phase. An exception is the +8.0~km\,s$^{-1}$ component toward NGC6334~I, in which HF is less than half as abundant as in other diffuse clouds. The chemical composition of this foreground component, however, seems to be different from the other components (e.g., strongly enhanced CH$^+$ abundance; Lis et al.~in prep.), which requires further studies.

In the high-mass star-forming regions, at much higher densities compared with diffuse clouds, the HF abundance seems to drop by about two orders of magnitude. The correlation with the physical conditions, especially density and temperature, would be crucial to determine the nature of this HF depletion, but it  still remains to be found. \cite{PBL10} argues that the low abundance of HF at higher densities is due to freeze out of HF onto dust grains. This argument is drawn from the fact that almost all fluorine should be bound in HF due to its high proton affinity, and that the desorption energy of HF is assumed to be quite large due to its polar nature. In dense ($\rm 10^{5}-10^6~cm^{-3}$) and warm  (100--150~K), but dynamically active regions such as the outflow in AFGL~2591, we find HF abundances close to the values found in diffuse clouds. In the picture of HF freeze out at high densities this can be explained by HF desorption due to sputtering by high velocity particles, similar to the  desorption of water in outflows (\citealt{KVv10}; \citealt{ELB10}). 

In both sources HF is the only fluorine bearing species we find in the HIFI line survey, and we can confirm non-detections for CF, CF$^+$, DF, and H$_2$F$^+$. In NGC6334~I the upper limit for CF$^+~J=5-4$, a molecule which has been detected in the Orion Bar with an  abundance of a few times $10^{-10}$ \citep{NSM06}, is 0.012~K. Assuming cloud properties, as determined by C$^{18}$O fits (Plume priv. communication), yields an upper limit of the CF$^+$ abundance of $6\cdot 10^{-12}$. This low upper limit, 50 times lower than the abundance found in the Orion Bar, can be explained by the different nature of the two objects. C$^+$, a precursor of CF$^+$, and thus CF$^+$ itself, are expected to be much more abundant in photo-dominated regions, such as the Orion Bar, than in massive cores. The upper limit of the abundance of these four fluorine bearing species together is $\sim 7\cdot 10^{-11}$, and thus these molecules hold less than 0.4\% of the total fluorine. The non-detection of other fluorine bearing species is another indication that freeze out onto dust grains causes the low HF abundances observed in dense, quiescent gas.

This analysis demonstrates that the utility of HF as a tracer of H$_2$, as predicted by chemical models, is only applicable to diffuse clouds. The fluorine chemistry in dense clouds is not yet fully understood and further theoretical and observational studies are required.

\acknowledgements

HIFI has been designed and built by a consortium of institutes and university departments from across Europe, Canada and the United States under the leadership of SRON Netherlands Institute for Space Research, Groningen, The Netherlands and with major contributions from Germany, France and the US. Consortium members are: Canada: CSA, U.Waterloo; France: CESR, LAB, LERMA, IRAM; Germany: KOSMA, MPIfR, MPS; Ireland, NUI Maynooth; Italy: ASI, IFSI-INAF, Osservatorio Astrofisico di Arcetri-INAF; Netherlands: SRON, TUD; Poland: CAMK, CBK; Spain: Observatorio Astron\'omico Nacional (IGN), Centro de Astrobiolog\'{\i}a (CSIC-INTA). Sweden: Chalmers University of Technology–MC2, RSS \& GARD; Onsala Space Observatory; Swedish National Space Board, Stockholm University – Stockholm Observatory; Switzerland: ETH Zurich, FHNW; USA: Caltech, JPL, NHSC. We thank many funding agencies for financial support.  Support for this work was provided by NASA through an award issued by JPL/Caltech, by l'Agence Nationale pour la Recherche (ANR), France (project FORCOMS, contracts ANR-08-BLAN-022) and the Centre National d’Etudes Spatiales (CNES).

\bibliographystyle{apj.bst}
\bibliography{library.bib}

\begin{thebibliography}{54}
\expandafter\ifx\csname natexlab\endcsname\relax\def\natexlab#1{#1}\fi

\bibitem[{{Ag{\'u}ndez} {et~al.}(2011){Ag{\'u}ndez}, {Cernicharo}, {Waters},
  {Decin}, {Encrenaz}, {Neufeld}, {Teyssier}, \& {Daniel}}]{ACW11}
{Ag{\'u}ndez}, M., {Cernicharo}, J., {Waters}, L.~B.~F.~M., {Decin}, L.,
  {Encrenaz}, P., {Neufeld}, D., {Teyssier}, D., \& {Daniel}, F. 2011, \aap,
  533, L6+

\bibitem[{{Beuther} {et~al.}(2007){Beuther}, {Walsh}, {Thorwirth}, {Zhang},
  {Hunter}, {Megeath}, \& {Menten}}]{BWT07}
{Beuther}, H., {Walsh}, A.~J., {Thorwirth}, S., {Zhang}, Q., {Hunter}, T.~R.,
  {Megeath}, S.~T., \& {Menten}, K.~M. 2007, \aap, 466, 989

\bibitem[{{Beuther} {et~al.}(2008){Beuther}, {Walsh}, {Thorwirth}, {Zhang},
  {Hunter}, {Megeath}, \& {Menten}}]{BWT08}
---. 2008, \aap, 481, 169

\bibitem[{{Bruderer} {et~al.}(2010){Bruderer}, {Benz}, {Van Dishoeck},
  {Melchior}, {Doty}, {Van der Tak}, {St{\"a}uber}, {Wampfler}, {Dedes},
  {Y{\i}ld{\i}z}, {Pagani}, {Giannini}, {de Graauw}, {Whyborn}, {Teyssier},
  {Jellema}, {Shipman}, {Schieder}, {Honingh}, {Caux}, {B{\"a}chtold},
  {Csillaghy}, {Monstein}, {Bachiller}, {Baudry}, {Benedettini}, {Bergin},
  {Bjerkeli}, {Blake}, {Bontemps}, {Braine}, {Caselli}, {Cernicharo},
  {Codella}, {Daniel}, {di Giorgio}, {Dominik}, {Encrenaz}, {Fich}, {Fuente},
  {Goicoechea}, {Helmich}, {Herczeg}, {Herpin}, {Hogerheijde}, {Jacq},
  {Johnstone}, {J{\o}rgensen}, {Kristensen}, {Larsson}, {Lis}, {Liseau},
  {Marseille}, {McCoey}, {Melnick}, {Neufeld}, {Nisini}, {Olberg}, {Parise},
  {Pearson}, {Plume}, {Risacher}, {Santiago-Garc{\'{\i}}a}, {Saraceno},
  {Shipman}, {Tafalla}, {Van Kempen}, {Visser}, \& {Wyrowski}}]{BBv10}
{Bruderer}, S., {et~al.} 2010, \aap, 521, L44+

\bibitem[{{Ceccarelli} {et~al.}(2010){Ceccarelli}, {Bacmann}, {Boogert},
  {Caux}, {Dominik}, {Lefloch}, {Lis}, {Schilke}, {Van der Tak}, {Caselli},
  {Cernicharo}, {Codella}, {Comito}, {Fuente}, {Baudry}, {Bell}, {Benedettini},
  {Bergin}, {Blake}, {Bottinelli}, {Cabrit}, {Castets}, {Coutens}, {Crimier},
  {Demyk}, {Encrenaz}, {Falgarone}, {Gerin}, {Goldsmith}, {Helmich},
  {Hennebelle}, {Henning}, {Herbst}, {Hily-Blant}, {Jacq}, {Kahane}, {Kama},
  {Klotz}, {Langer}, {Lord}, {Lorenzani}, {Maret}, {Melnick}, {Neufeld},
  {Nisini}, {Pacheco}, {Pagani}, {Parise}, {Pearson}, {Phillips}, {Salez},
  {Saraceno}, {Schuster}, {Tielens}, {Van der Wiel}, {Vastel}, {Viti},
  {Wakelam}, {Walters}, {Wyrowski}, {Yorke}, {Liseau}, {Olberg}, {Szczerba},
  {Benz}, \& {Melchior}}]{CBB10}
{Ceccarelli}, C., {et~al.} 2010, \aap, 521, L22+

\bibitem[{{Choi} {et~al.}(2011){Choi}, {Van der Tak}, \& {Van
  Dishoeck}}]{Cvv11}
{Choi}, Y., {Van der Tak}, F., \& {Van Dishoeck}, E.~F. 2011, in IAU Symposium,
  Vol. 280, IAU Symposium, 128P--+

\bibitem[{{Comito} \& {Schilke}(2002)}]{CS02}
{Comito}, C., \& {Schilke}, P. 2002, \aap, 395, 357

\bibitem[{{de Graauw} {et~al.}(2010){de Graauw}, {Helmich}, {Phillips},
  {Stutzki}, {Caux}, {Whyborn}, {Dieleman}, {Roelfsema}, {Aarts}, {Assendorp},
  {Bachiller}, {Baechtold}, {Barcia}, {Beintema}, {Belitsky}, {Benz}, {Bieber},
  {Boogert}, {Borys}, {Bumble}, {Ca{\"i}s}, {Caris}, {Cerulli-Irelli},
  {Chattopadhyay}, {Cherednichenko}, {Ciechanowicz}, {Coeur-Joly}, {Comito},
  {Cros}, {de Jonge}, {de Lange}, {Delforges}, {Delorme}, {den Boggende},
  {Desbat}, {Diez-Gonz{\'a}lez}, {di Giorgio}, {Dubbeldam}, {Edwards},
  {Eggens}, {Erickson}, {Evers}, {Fich}, {Finn}, {Franke}, {Gaier}, {Gal},
  {Gao}, {Gallego}, {Gauffre}, {Gill}, {Glenz}, {Golstein}, {Goulooze},
  {Gunsing}, {G{\"u}sten}, {Hartogh}, {Hatch}, {Higgins}, {Honingh}, {Huisman},
  {Jackson}, {Jacobs}, {Jacobs}, {Jarchow}, {Javadi}, {Jellema}, {Justen},
  {Karpov}, {Kasemann}, {Kawamura}, {Keizer}, {Kester}, {Klapwijk}, {Klein},
  {Kollberg}, {Kooi}, {Kooiman}, {Kopf}, {Krause}, {Krieg}, {Kramer},
  {Kruizenga}, {Kuhn}, {Laauwen}, {Lai}, {Larsson}, {Leduc}, {Leinz}, {Lin},
  {Liseau}, {Liu}, {Loose}, {L{\'o}pez-Fernandez}, {Lord}, {Luinge}, {Marston},
  {Mart{\'{\i}}n-Pintado}, {Maestrini}, {Maiwald}, {McCoey}, {Mehdi}, {Megej},
  {Melchior}, {Meinsma}, {Merkel}, {Michalska}, {Monstein}, {Moratschke},
  {Morris}, {Muller}, {Murphy}, {Naber}, {Natale}, {Nowosielski}, {Nuzzolo},
  {Olberg}, {Olbrich}, {Orfei}, {Orleanski}, {Ossenkopf}, {Peacock}, {Pearson},
  {Peron}, {Phillip-May}, {Piazzo}, {Planesas}, {Rataj}, {Ravera}, {Risacher},
  {Salez}, {Samoska}, {Saraceno}, {Schieder}, {Schlecht}, {Schl{\"o}der},
  {Schm{\"u}lling}, {Schultz}, {Schuster}, {Siebertz}, {Smit}, {Szczerba},
  {Shipman}, {Steinmetz}, {Stern}, {Stokroos}, {Teipen}, {Teyssier}, {Tils},
  {Trappe}, {Van Baaren}, {Van Leeuwen}, {Van de Stadt}, {Visser}, {Wildeman},
  {Wafelbakker}, {Ward}, {Wesselius}, {Wild}, {Wulff}, {Wunsch}, {Tielens},
  {Zaal}, {Zirath}, {Zmuidzinas}, \& {Zwart}}]{dHP10}
{de Graauw}, T., {et~al.} 2010, \aap, 518, L6+

\bibitem[{{de Pree} {et~al.}(1995){de Pree}, {Rodriguez}, {Dickel}, \&
  {Goss}}]{dRD95}
{de Pree}, C.~G., {Rodriguez}, L.~F., {Dickel}, H.~R., \& {Goss}, W.~M. 1995,
  \apj, 447, 220

\bibitem[{{de Wit} {et~al.}(2009){de Wit}, {Hoare}, {Fujiyoshi}, {Oudmaijer},
  {Honda}, {Kataza}, {Miyata}, {Okamoto}, {Onaka}, {Sako}, \&
  {Yamashita}}]{dHF09}
{de Wit}, W.~J., {et~al.} 2009, \aap, 494, 157

\bibitem[{{Doty} {et~al.}(2002){Doty}, {Van Dishoeck}, {Van der Tak}, \&
  {Boonman}}]{Dvv02}
{Doty}, S.~D., {Van Dishoeck}, E.~F., {Van der Tak}, F.~F.~S., \& {Boonman},
  A.~M.~S. 2002, \aap, 389, 446

\bibitem[{{Ellingsen} {et~al.}(1996){Ellingsen}, {von Bibra}, {McCulloch},
  {Norris}, {Deshpande}, \& {Phillips}}]{EvM96}
{Ellingsen}, S.~P., {von Bibra}, M.~L., {McCulloch}, P.~M., {Norris}, R.~P.,
  {Deshpande}, A.~A., \& {Phillips}, C.~J. 1996, \mnras, 280, 378

\bibitem[{{Emprechtinger} {et~al.}(2010){Emprechtinger}, {Lis}, {Bell},
  {Phillips}, {Schilke}, {Comito}, {Rolffs}, {Van der Tak}, {Ceccarelli},
  {Aarts}, {Bacmann}, {Baudry}, {Benedettini}, {Bergin}, {Blake}, {Boogert},
  {Bottinelli}, {Cabrit}, {Caselli}, {Castets}, {Caux}, {Cernicharo},
  {Codella}, {Coutens}, {Crimier}, {Demyk}, {Dominik}, {Encrenaz}, {Falgarone},
  {Fuente}, {Gerin}, {Goldsmith}, {Helmich}, {Hennebelle}, {Henning}, {Herbst},
  {Hily-Blant}, {Jacq}, {Kahane}, {Kama}, {Klotz}, {Kooi}, {Langer}, {Lefloch},
  {Loose}, {Lord}, {Lorenzani}, {Maret}, {Melnick}, {Neufeld}, {Nisini},
  {Ossenkopf}, {Pacheco}, {Pagani}, {Parise}, {Pearson}, {Risacher}, {Salez},
  {Saraceno}, {Schuster}, {Stutzki}, {Tielens}, {Van der Wiel}, {Vastel},
  {Viti}, {Wakelam}, {Walters}, {Wyrowski}, \& {Yorke}}]{ELB10}
{Emprechtinger}, M., {et~al.} 2010, \aap, 521, L28+

\bibitem[{{Flower} \& {Pineau Des For{\^e}ts}(2010)}]{FP10}
{Flower}, D.~R., \& {Pineau Des For{\^e}ts}, G. 2010, \mnras, 406, 1745

\bibitem[{{Godard} {et~al.}(2009){Godard}, {Falgarone}, \& {Pineau Des
  For{\^e}ts}}]{GFP09}
{Godard}, B., {Falgarone}, E., \& {Pineau Des For{\^e}ts}, G. 2009, \aap, 495,
  847

\bibitem[{{Guillon} {et~al.}(2011){Guillon}, {Benz}, {Van Dishoeck},
  {Melchior}, {Doty}, {Van der Tak}, {St{\"a}uber}, {Wampfler}, {Dedes},
  {Y{\i}ld{\i}z}, {Pagani}, {Giannini}, {de Graauw}, {Whyborn}, {Teyssier},
  {Jellema}, {Shipman}, {Schieder}, {Honingh}, {Caux}, {B{\"a}chtold},
  {Csillaghy}, {Monstein}, {Bachiller}, {Baudry}, {Benedettini}, {Bergin},
  {Bjerkeli}, {Blake}, {Bontemps}, {Braine}, {Caselli}, {Cernicharo},
  {Codella}, {Daniel}, {di Giorgio}, {Dominik}, {Encrenaz}, {Fich}, {Fuente},
  {Goicoechea}, {Helmich}, {Herczeg}, {Herpin}, {Hogerheijde}, {Jacq},
  {Johnstone}, {J{\o}rgensen}, {Kristensen}, {Larsson}, {Lis}, {Liseau},
  {Marseille}, {McCoey}, {Melnick}, {Neufeld}, {Nisini}, {Olberg}, {Parise},
  {Pearson}, {Plume}, {Risacher}, {Santiago-Garc{\'{\i}}a}, {Saraceno},
  {Shipman}, {Tafalla}, {Van Kempen}, {Visser}, \& {Wyrowski}}]{Gprep}
{Guillon}, G., {et~al.} 2011, MNRAS, accepted

\bibitem[{{Hasegawa} \& {Mitchell}(1995)}]{HM95}
{Hasegawa}, T.~I., \& {Mitchell}, G.~F. 1995, \apj, 451, 225

\bibitem[{{Hogerheijde} \& {Van der Tak}(2000)}]{Hv00}
{Hogerheijde}, M.~R., \& {Van der Tak}, F.~F.~S. 2000, \aap, 362, 697

\bibitem[{{Hunter} {et~al.}(2006){Hunter}, {Brogan}, {Megeath}, {Menten},
  {Beuther}, \& {Thorwirth}}]{HBM06}
{Hunter}, T.~R., {Brogan}, C.~L., {Megeath}, S.~T., {Menten}, K.~M., {Beuther},
  H., \& {Thorwirth}, S. 2006, \apj, 649, 888

\bibitem[{{Kraemer} \& {Jackson}(1995)}]{KJ95}
{Kraemer}, K.~E., \& {Jackson}, J.~M. 1995, \apjl, 439, L9

\bibitem[{{Kristensen} {et~al.}(2010){Kristensen}, {Visser}, {Van Dishoeck},
  {Y{\i}ld{\i}z}, {Doty}, {Herczeg}, {Liu}, {Parise}, {J{\o}rgensen}, {Van
  Kempen}, {Brinch}, {Wampfler}, {Bruderer}, {Benz}, {Hogerheijde}, {Deul},
  {Bachiller}, {Baudry}, {Benedettini}, {Bergin}, {Bjerkeli}, {Blake},
  {Bontemps}, {Braine}, {Caselli}, {Cernicharo}, {Codella}, {Daniel}, {de
  Graauw}, {di Giorgio}, {Dominik}, {Encrenaz}, {Fich}, {Fuente}, {Giannini},
  {Goicoechea}, {Helmich}, {Herpin}, {Jacq}, {Johnstone}, {Kaufman}, {Larsson},
  {Lis}, {Liseau}, {Marseille}, {McCoey}, {Melnick}, {Neufeld}, {Nisini},
  {Olberg}, {Pearson}, {Plume}, {Risacher}, {Santiago-Garc{\'{\i}}a},
  {Saraceno}, {Shipman}, {Tafalla}, {Tielens}, {Van der Tak}, {Wyrowski},
  {Beintema}, {de Jonge}, {Dieleman}, {Ossenkopf}, {Roelfsema}, {Stutzki}, \&
  {Whyborn}}]{KVv10}
{Kristensen}, L.~E., {et~al.} 2010, \aap, 521, L30+

\bibitem[{{Lada} {et~al.}(1984){Lada}, {Thronson}, {Smith}, {Schwartz}, \&
  {Glaccum}}]{LTS84}
{Lada}, C.~J., {Thronson}, Jr., H.~A., {Smith}, H.~A., {Schwartz}, P.~R., \&
  {Glaccum}, W. 1984, \apj, 286, 302

\bibitem[{{Leurini} {et~al.}(2006){Leurini}, {Schilke}, {Parise}, {Wyrowski},
  {G{\"u}sten}, \& {Philipp}}]{LSP06}
{Leurini}, S., {Schilke}, P., {Parise}, B., {Wyrowski}, F., {G{\"u}sten}, R.,
  \& {Philipp}, S. 2006, \aap, 454, L83

\bibitem[{{Lis} {et~al.}(2010){Lis}, {Pearson}, {Neufeld}, {Schilke},
  {M{\"u}ller}, {Gupta}, {Bell}, {Comito}, {Phillips}, {Bergin}, {Ceccarelli},
  {Goldsmith}, {Blake}, {Bacmann}, {Baudry}, {Benedettini}, {Benz}, {Black},
  {Boogert}, {Bottinelli}, {Cabrit}, {Caselli}, {Castets}, {Caux},
  {Cernicharo}, {Codella}, {Coutens}, {Crimier}, {Crockett}, {Daniel}, {Demyk},
  {Dominic}, {Dubernet}, {Emprechtinger}, {Encrenaz}, {Falgarone}, {Fuente},
  {Gerin}, {Giesen}, {Goicoechea}, {Helmich}, {Hennebelle}, {Henning},
  {Herbst}, {Hily-Blant}, {Hjalmarson}, {Hollenbach}, {Jack}, {Joblin},
  {Johnstone}, {Kahane}, {Kama}, {Kaufman}, {Klotz}, {Langer}, {Larsson}, {Le
  Bourlot}, {Lefloch}, {Le Petit}, {Li}, {Liseau}, {Lord}, {Lorenzani},
  {Maret}, {Martin}, {Melnick}, {Menten}, {Morris}, {Murphy}, {Nagy}, {Nisini},
  {Ossenkopf}, {Pacheco}, {Pagani}, {Parise}, {P{\'e}rault}, {Plume}, {Qin},
  {Roueff}, {Salez}, {Sandqvist}, {Saraceno}, {Schlemmer}, {Schuster}, {Snell},
  {Stutzki}, {Tielens}, {Trappe}, {Van der Tak}, {Van der Wiel}, {Van
  Dishoeck}, {Vastel}, {Viti}, {Wakelam}, {Walters}, {Wang}, {Wyrowski},
  {Yorke}, {Yu}, {Zmuidzinas}, {Delorme}, {Desbat}, {G{\"u}sten}, {Krieg}, \&
  {Delforge}}]{LPN10}
{Lis}, D.~C., {et~al.} 2010, \aap, 521, L9+

\bibitem[{{Monje} {et~al.}(2011){Monje}, {Emprechtinger}, {Phillips}, {Lis},
  {Goldsmith}, {Bergin}, {Bell}, {Neufeld}, \& {Sonnentrucker}}]{MEP11}
{Monje}, R.~R., {et~al.} 2011, \apjl, 734, L23+

\bibitem[{{M{\"u}ller} {et~al.}(2005){M{\"u}ller}, {Schl{\"o}der}, {Stutzki},
  \& {Winnewisser}}]{MSS05}
{M{\"u}ller}, H.~S.~P., {Schl{\"o}der}, F., {Stutzki}, J., \& {Winnewisser}, G.
  2005, Journal of Molecular Structure, 742, 215

\bibitem[{{Neckel}(1978)}]{N78}
{Neckel}, T. 1978, \aap, 69, 51

\bibitem[{{Neufeld} {et~al.}(2005){Neufeld}, {Wolfire}, \& {Schilke}}]{NWS05}
{Neufeld}, D.~A., {Wolfire}, M.~G., \& {Schilke}, P. 2005, \apj, 628, 260

\bibitem[{{Neufeld} {et~al.}(1997){Neufeld}, {Zmuidzinas}, {Schilke}, \&
  {Phillips}}]{NZS97}
{Neufeld}, D.~A., {Zmuidzinas}, J., {Schilke}, P., \& {Phillips}, T.~G. 1997,
  \apjl, 488, L141+

\bibitem[{{Neufeld} {et~al.}(2006){Neufeld}, {Schilke}, {Menten}, {Wolfire},
  {Black}, {Schuller}, {M{\"u}ller}, {Thorwirth}, {G{\"u}sten}, \&
  {Philipp}}]{NSM06}
{Neufeld}, D.~A., {et~al.} 2006, \aap, 454, L37

\bibitem[{{Neufeld} {et~al.}(2010){Neufeld}, {Sonnentrucker}, {Phillips},
  {Lis}, {de Luca}, {Goicoechea}, {Black}, {Gerin}, {Bell}, {Boulanger},
  {Cernicharo}, {Coutens}, {Dartois}, {Kazmierczak}, {Encrenaz}, {Falgarone},
  {Geballe}, {Giesen}, {Godard}, {Goldsmith}, {Gry}, {Gupta}, {Hennebelle},
  {Herbst}, {Hily-Blant}, {Joblin}, {Ko{\l}os}, {Kre{\l}owski},
  {Mart{\'{\i}}n-Pintado}, {Menten}, {Monje}, {Mookerjea}, {Pearson},
  {Perault}, {Persson}, {Plume}, {Salez}, {Schlemmer}, {Schmidt}, {Stutzki},
  {Teyssier}, {Vastel}, {Yu}, {Cais}, {Caux}, {Liseau}, {Morris}, \&
  {Planesas}}]{NSP10}
---. 2010, \aap, 518, L108+

\bibitem[{{Ossenkopf} {et~al.}(2010){Ossenkopf}, {M{\"u}ller}, {Lis},
  {Schilke}, {Bell}, {Bruderer}, {Bergin}, {Ceccarelli}, {Comito}, {Stutzki},
  {Bacman}, {Baudry}, {Benz}, {Benedettini}, {Berne}, {Blake}, {Boogert},
  {Bottinelli}, {Boulanger}, {Cabrit}, {Caselli}, {Caux}, {Cernicharo},
  {Codella}, {Coutens}, {Crimier}, {Crockett}, {Daniel}, {Demyk}, {Dieleman},
  {Dominik}, {Dubernet}, {Emprechtinger}, {Encrenaz}, {Falgarone}, {France},
  {Fuente}, {Gerin}, {Giesen}, {di Giorgio}, {Goicoechea}, {Goldsmith},
  {G{\"u}sten}, {Harris}, {Helmich}, {Herbst}, {Hily-Blant}, {Jacobs}, {Jacq},
  {Joblin}, {Johnstone}, {Kahane}, {Kama}, {Klein}, {Klotz}, {Kramer},
  {Langer}, {Lefloch}, {Leinz}, {Lorenzani}, {Lord}, {Maret}, {Martin},
  {Martin-Pintado}, {McCoey}, {Melchior}, {Melnick}, {Menten}, {Mookerjea},
  {Morris}, {Murphy}, {Neufeld}, {Nisini}, {Pacheco}, {Pagani}, {Parise},
  {Pearson}, {P{\'e}rault}, {Phillips}, {Plume}, {Quin}, {Rizzo}, {R{\"o}llig},
  {Salez}, {Saraceno}, {Schlemmer}, {Simon}, {Schuster}, {Van der Tak},
  {Tielens}, {Teyssier}, {Trappe}, {Vastel}, {Viti}, {Wakelam}, {Walters},
  {Wang}, {Whyborn}, {Van der Wiel}, {Yorke}, {Yu}, \& {Zmuidzinas}}]{OML10}
{Ossenkopf}, V., {et~al.} 2010, \aap, 518, L111+

\bibitem[{{Ott}(2010)}]{O10}
{Ott}, S. 2010, in Astronomical Society of the Pacific Conference Series, Vol.
  434, Astronomical Data Analysis Software and Systems XIX, ed. {Y.~Mizumoto,
  K.-I.~Morita, \& M.~Ohishi}, 139--+

\bibitem[{{Phillips} {et~al.}(2010){Phillips}, {Bergin}, {Lis}, {Neufeld},
  {Bell}, {Wang}, {Crockett}, {Emprechtinger}, {Blake}, {Caux}, {Ceccarelli},
  {Cernicharo}, {Comito}, {Daniel}, {Dubernet}, {Encrenaz}, {Gerin}, {Giesen},
  {Goicoechea}, {Goldsmith}, {Herbst}, {Joblin}, {Johnstone}, {Langer},
  {Latter}, {Lord}, {Maret}, {Martin}, {Melnick}, {Menten}, {Morris},
  {M{\"u}ller}, {Murphy}, {Ossenkopf}, {Pearson}, {P{\'e}rault}, {Plume},
  {Qin}, {Schilke}, {Schlemmer}, {Stutzki}, {Trappe}, {Van der Tak}, {Vastel},
  {Yorke}, {Yu}, {Zmuidzinas}, {Boogert}, {G{\"u}sten}, {Hartogh}, {Honingh},
  {Karpov}, {Kooi}, {Krieg}, \& {Schieder}}]{PBL10}
{Phillips}, T.~G., {et~al.} 2010, \aap, 518, L109+

\bibitem[{{Pilbratt} {et~al.}(2010){Pilbratt}, {Riedinger}, {Passvogel},
  {Crone}, {Doyle}, {Gageur}, {Heras}, {Jewell}, {Metcalfe}, {Ott}, \&
  {Schmidt}}]{PRP10}
{Pilbratt}, G.~L., {et~al.} 2010, \aap, 518, L1+

\bibitem[{{Rolffs} {et~al.}(2011){Rolffs}, {Schilke}, {Wyrowski}, {Menten},
  {G{\"u}sten}, \& {Bisschop}}]{RSW11}
{Rolffs}, R., {Schilke}, P., {Wyrowski}, F., {Menten}, K.~M., {G{\"u}sten}, R.,
  \& {Bisschop}, S.~E. 2011, \aap, 527, A68+

\bibitem[{{Rygl} {et~al.}(2012){Rygl}, {Brunthaler}, {Sanna}, {Menten}, {Reid},
  {van Langevelde}, {Honma}, {Torstensson}, \& {Fujisawa}}]{RBS12}
{Rygl}, K.~L.~J., {et~al.} 2012, \aap, 539, A79

\bibitem[{{Sandell}(2000)}]{S00}
{Sandell}, G. 2000, \aap, 358, 242

\bibitem[{{Schilke} {et~al.}(2006){Schilke}, {Comito}, {Thorwirth}, {Wyrowski},
  {Menten}, {G{\"u}sten}, {Bergman}, \& {Nyman}}]{SCT06}
{Schilke}, P., {Comito}, C., {Thorwirth}, S., {Wyrowski}, F., {Menten}, K.~M.,
  {G{\"u}sten}, R., {Bergman}, P., \& {Nyman}, L.-{\AA}. 2006, \aap, 454, L41

\bibitem[{{Sch{\"o}ier} {et~al.}(2005){Sch{\"o}ier}, {Van der Tak}, {Van
  Dishoeck}, \& {Black}}]{Svv05}
{Sch{\"o}ier}, F.~L., {Van der Tak}, F.~F.~S., {Van Dishoeck}, E.~F., \&
  {Black}, J.~H. 2005, \aap, 432, 369

\bibitem[{{Sheffer} {et~al.}(2008){Sheffer}, {Rogers}, {Federman}, {Abel},
  {Gredel}, {Lambert}, \& {Shaw}}]{SRF08}
{Sheffer}, Y., {Rogers}, M., {Federman}, S.~R., {Abel}, N.~P., {Gredel}, R.,
  {Lambert}, D.~L., \& {Shaw}, G. 2008, \apj, 687, 1075

\bibitem[{{Sonnentrucker} {et~al.}(2007){Sonnentrucker}, {Welty}, {Thorburn},
  \& {York}}]{SWT07}
{Sonnentrucker}, P., {Welty}, D.~E., {Thorburn}, J.~A., \& {York}, D.~G. 2007,
  \apjs, 168, 58

\bibitem[{{Sonnentrucker} {et~al.}(2010){Sonnentrucker}, {Neufeld}, {Phillips},
  {Gerin}, {Lis}, {de Luca}, {Goicoechea}, {Black}, {Bell}, {Boulanger},
  {Cernicharo}, {Coutens}, {Dartois}, {Ka{\'z}mierczak}, {Encrenaz},
  {Falgarone}, {Geballe}, {Giesen}, {Godard}, {Goldsmith}, {Gry}, {Gupta},
  {Hennebelle}, {Herbst}, {Hily-Blant}, {Joblin}, {Ko{\l}os}, {Kre{\l}owski},
  {Mart{\'{\i}}n-Pintado}, {Menten}, {Monje}, {Mookerjea}, {Pearson},
  {Perault}, {Persson}, {Plume}, {Salez}, {Schlemmer}, {Schmidt}, {Stutzki},
  {Teyssier}, {Vastel}, {Yu}, {Caux}, {G{\"u}sten}, {Hatch}, {Klein}, {Mehdi},
  {Morris}, \& {Ward}}]{SNP10}
{Sonnentrucker}, P., {et~al.} 2010, \aap, 521, L12+

\bibitem[{{Straw} \& {Hyland}(1989)}]{SH89}
{Straw}, S.~M., \& {Hyland}, A.~R. 1989, \apj, 340, 318

\bibitem[{{Thorwirth} {et~al.}(2003){Thorwirth}, {Winnewisser}, {Megeath}, \&
  {Tieftrunk}}]{TWM03}
{Thorwirth}, S., {Winnewisser}, G., {Megeath}, S.~T., \& {Tieftrunk}, A.~R.
  2003, in Astronomical Society of the Pacific Conference Series, Vol. 287,
  Galactic Star Formation Across the Stellar Mass Spectrum, ed. {J.~M.~De
  Buizer \& N.~S.~Van der Bliek}, 257--260

\bibitem[{{Van der Tak} \& {Menten}(2005)}]{vM05}
{Van der Tak}, F.~F.~S., \& {Menten}, K.~M. 2005, \aap, 437, 947

\bibitem[{{Van der Tak} {et~al.}(2012){Van der Tak}, {Ossenkopf}, {Nagy},
  {Faure}, {R{\"o}llig}, \& {Bergin}}]{vON12}
{Van der Tak}, F.~F.~S., {Ossenkopf}, V., {Nagy}, Z., {Faure}, A.,
  {R{\"o}llig}, M., \& {Bergin}, E.~A. 2012, \aap, 537, L10

\bibitem[{{Van der Tak} {et~al.}(1999){Van der Tak}, {Van Dishoeck}, {Evans},
  {Bakker}, \& {Blake}}]{vvE99}
{Van der Tak}, F.~F.~S., {Van Dishoeck}, E.~F., {Evans}, II, N.~J., {Bakker},
  E.~J., \& {Blake}, G.~A. 1999, \apj, 522, 991

\bibitem[{{Van der Tak} {et~al.}(2000){Van der Tak}, {Van Dishoeck}, {Evans},
  \& {Blake}}]{vvE00}
{Van der Tak}, F.~F.~S., {Van Dishoeck}, E.~F., {Evans}, II, N.~J., \& {Blake},
  G.~A. 2000, \apj, 537, 283

\bibitem[{{Van der Werf} {et~al.}(2010){Van der Werf}, {Isaak}, {Meijerink},
  {Spaans}, {Rykala}, {Fulton}, {Loenen}, {Walter}, {Wei{\ss}}, {Armus},
  {Fischer}, {Israel}, {Harris}, {Veilleux}, {Henkel}, {Savini}, {Lord},
  {Smith}, {Gonz{\'a}lez-Alfonso}, {Naylor}, {Aalto}, {Charmandaris}, {Dasyra},
  {Evans}, {Gao}, {Greve}, {G{\"u}sten}, {Kramer}, {Mart{\'{\i}}n-Pintado},
  {Mazzarella}, {Papadopoulos}, {Sanders}, {Spinoglio}, {Stacey}, {Vlahakis},
  {Wiedner}, \& {Xilouris}}]{vIM10}
{Van der Werf}, P.~P., {et~al.} 2010, \aap, 518, L42+

\bibitem[{{Van der Wiel}(2011)}]{v11}
{Van der Wiel}, M.~H.~D. 2011, "Molecular gas and dust influenced by massive
  protostars : spectral surveys in the far-infrared and submillimeter",
  PhD-Thesis, University of Groningen

\bibitem[{{Van der Wiel} {et~al.}(2011){Van der Wiel}, {Van der Tak}, {Spaans},
  {Fuller}, {Plume}, {Roberts}, \& {Williams}}]{vvS11}
{Van der Wiel}, M.~H.~D., {Van der Tak}, F.~F.~S., {Spaans}, M., {Fuller},
  G.~A., {Plume}, R., {Roberts}, H., \& {Williams}, J.~L. 2011, \aap, 532, A88+

\bibitem[{{Van der Wiel} {et~al.}(2010){Van der Wiel}, {Van der Tak}, {Lis},
  {Bell}, {Bergin}, {Comito}, {Emprechtinger}, {Schilke}, {Caux}, {Ceccarelli},
  {Baudry}, {Goldsmith}, {Herbst}, {Langer}, {Lord}, {Neufeld}, {Pearson},
  {Phillips}, {Rolffs}, {Yorke}, {Bacmann}, {Benedettini}, {Blake}, {Boogert},
  {Bottinelli}, {Cabrit}, {Caselli}, {Castets}, {Cernicharo}, {Codella},
  {Coutens}, {Crimier}, {Demyk}, {Dominik}, {Encrenaz}, {Falgarone}, {Fuente},
  {Gerin}, {Helmich}, {Hennebelle}, {Henning}, {Hily-Blant}, {Jacq}, {Kahane},
  {Kama}, {Klotz}, {Lefloch}, {Lorenzani}, {Maret}, {Melnick}, {Nisini},
  {Pacheco}, {Pagani}, {Parise}, {Salez}, {Saraceno}, {Schuster}, {Tielens},
  {Vastel}, {Viti}, {Wakelam}, {Walters}, {Wyrowski}, {Edwards}, {Zmuidzinas},
  {Morris}, {Samoska}, \& {Teyssier}}]{vvL10}
{Van der Wiel}, M.~H.~D., {et~al.} 2010, \aap, 521, L43+

\bibitem[{{Walsh} {et~al.}(2007){Walsh}, {Longmore}, {Thorwirth}, {Urquhart},
  \& {Purcell}}]{WLT07}
{Walsh}, A.~J., {Longmore}, S.~N., {Thorwirth}, S., {Urquhart}, J.~S., \&
  {Purcell}, C.~R. 2007, \mnras, 382, L35

\end{thebibliography}
\end{document}